\documentclass{article} 
\usepackage{iclr2020_conference,times}

\usepackage{hyperref}
\usepackage{url}
\usepackage[utf8]{inputenc}
\usepackage[russian, english]{babel}
\usepackage{fixltx2e}
\usepackage[ruled,vlined]{algorithm2e}
\usepackage{graphicx}
\usepackage{subfigure}
\usepackage{floatrow}
\newfloatcommand{capbtabbox}{table}[][\FBwidth]

\usepackage{blindtext}

\title{Attacking Neural Text Detectors}

\author{Max Wolff\thanks{No longer at Viewpoint School. Please contact at \texttt{m.wolff1621@gmail.com}.}  \\
Viewpoint School\\
Calabasas, CA 91302, USA \\
\texttt{m.wolff20@viewpoint.org}
\And
Stuart Wolff
}

\iclrfinalcopy 
\begin{document}

\maketitle

\begin{abstract}

Machine learning based language models have recently made significant progress, which introduces a danger to spread misinformation. To combat this potential danger, several methods have been proposed for detecting text written by these language models. This paper presents two classes of black-box attacks on these detectors, one which randomly replaces characters with homoglyphs, and the other a simple scheme to purposefully misspell words. The homoglyph and misspelling attacks decrease a popular neural text detector's recall on neural text from 97.44\% to 0.26\% and 22.68\%, respectively. Results also indicate that the attacks are transferable to other neural text detectors. 

\end{abstract}

\section{Introduction}
\label{intro}


Contemporary state of the art language models such as GPT-2 \citep{gpt-2} are rapidly improving, as they are being trained on increasingly large datasets and defined using billions of parameters. Language models are currently able to generate coherent text that humans can identify as machine-written text (neural text) with approximately 54\% accuracy. \citep{gltr}--close to random guessing. With this increasing power, language models provide bad actors with the potential to spread misinformation on an unprecedented scale \citep{gpt-2_report} and undermine clear authorship. 

To reduce the spread of misinformation via language models and give readers a better sense of what entity (machine or human) may have actually written a piece of text, multiple neural text detection methods have been proposed. Two automatic neural text detectors are considered in this work, RoBERTa \citep{gpt-2_report, roberta} and GROVER \citep{grover}, which are 95\% and 92\% accurate in discriminating neural text from human-written text, respectively.
Another tool, GLTR \citep{gltr}, is designed to assist humans in detecting neural text, increasing humans' ability to correctly distinguish neural text and human-written text from 54\% to 72\%. Fundamentally, these detectors are based on the fact that neural text follows predictable patterns based on the neural text's underlying language model generator.

Attacks on machine learning models, called adversarial attacks \citep{props, robust, glasses, hotflip}, have been studied in depth and used to expose both security holes and understand how machine learning models function by purposefully causing machine learning models to make mistakes.


Historically, homoglyph attacks\footnote{\url{https://en.wikipedia.org/wiki/IDN_homograph_attack}} have been used to direct victims to malicious websites by replacing characters in a trusted URL with similar looking ones, called homoglyphs. Part of this work seeks to test whether homoglyph attacks can also be used to create effective black-box adversarial attacks on neural text detectors.


\section{Threat Model and Proposed Attacks}

In this paper, two classes of attacks on neural text detectors are proposed. Both of these attacks attempt to modify neural text in ways that are relatively visually imperceptible to humans, but will cause a neural text detector to misclassify the text as human-written. Specifically, these attacks change the underlying distribution of neural text so that it diverges from that of the language model which generated it.

The first class of attacks are non human-like attacks, which imperceptibly (according to humans) change neural text in a way that humans normally would not. This class of attack shifts the modified text's distribution away from its original one. In this work, the non-human like attacks are realized by swapping selected characters with Unicode homoglyphs (e.g. changing English ``a''s to Cyrillic ``a''s throughout a neural text sample). Homoglyphs are chosen because they appear visually similar to their counterparts, but get tokenized differently by neural text detectors.


The second class of attacks are human-like attacks, which imperceptibly (according to humans) change neural text in a way that humans normally would. In this paper, this class of attack is realized by randomly swapping correctly spelled words with common human misspellings throughout a neural next sample--which from here onward is referred to as a ``misspelling attack.'' However, this is not the only way human-like attacks may be implemented. This class of attack may also target word-choice, grammar, or punctuation. Misspelling attacks are simply a proof-of-concept for this larger umbrella of human-like attacks.

\section{Experiments}
\label{methods}

A neural text dataset containing 5,000 text samples generated by GPT-2 1.5B using top-k 40 sampling was used to evaluate attacks in all experiments. This dataset was taken from a GitHub repository.\footnote{\url{https://github.com/openai/gpt-2-output-dataset}}
 In all experiments, except for the transferability tests, an open source implementation of the automatic RoBERTa neural text detector\footnote{\url{https://github.com/openai/gpt-2-output-dataset/tree/master/detector}} was used. Before the attacks, RoBERTa's recall on neural text was 97.44\%. In this paper, five experiments testing homoglyph attacks were conducted, and two were conducted for misspelling attacks.

The first homoglyph experiment in this paper was designed to test the effectiveness of different homoglyph pairs in lowering detector recall on neural text. In this experiment, all attacks were restricted to randomly replacing 1.5\% of all the characters in a given neural text sample to homoglyphs. If there were not enough of the character(s) being replaced in a neural text sample to meet this 1.5\% quota, the text sample was thrown out and the result of the attack not considered. Even so, every attack in experiments conducted under these conditions was run on at least 2,500 neural text samples. 

The second homoglyph experiment took the most effective homoglyph pair found in the first experiment and tested the effectiveness of the homoglyph attack when it was allowed to replace every occurrence of the target character(s). 

The third homoglyph experiment was designed to take the most effective homoglyph pair and test how varying frequencies of replacement may affect detector recall on neural text.

The fourth homoglyph experiment was designed to test the transferability of the homoglyph attacks to the GROVER and GLTR online demos. \footnote{\url{https://grover.allenai.org/detect}} \footnote{\url{http://gltr.io/dist/index.html}} In this experiment, 20 samples of neural text were randomly selected from the neural text dataset. Then, the most effective homoglyph attack (found in the first homoglyph experiment) was applied to the samples. GROVER's predictions on the original neural text and modified neural text were then recorded. The online demo for GROVER outputs ``We are quite sure this was written by a machine,'' (Machine++) ``We think this was written by a machine (but we're not sure),''(Machine+) ``We think this was written by a human (but we're not sure),'' (Human+) or ``We are quite sure this was written by a human'' (Human++). A similar experiment was performed on the GLTR demo. The most successful homoglyph attack was applied to 10 samples of text taken randomly from the neural text dataset.\footnote{The GLTR interface does not allow many Unicode characters, including Cyrillic ones. Thus, the homoglyph attack used for the GLTR experiments was the most successful, GLTR-allowed homoglyph attack.} Screenshots of GLTR's graphical interface were then taken before and after the attack, and patterns were observed. 

For the misspelling attack experiments, words were randomly misspelled throughout a text sample using a Wikipedia list\footnote{\url{https://en.wikipedia.org/wiki/Wikipedia:Lists_of_common_misspellings/For_machines}} of commonly misspelled words (by humans) in the English language. The attack was restricted to randomly misspelling 5\% of the words in each neural text sample in the dataset. The same transferability experiments used in the homoglyph attacks were used for the misspelling attacks, except instead of characters being replaced with homoglyphs, a random 5\% of the words in neural text samples were misspelled.

Code to reproduce results found in this paper can be found at \url{https://github.com/mwolff31/attacking_neural_text_detectors}.

\begin{table}[t]
\caption{\centering RoBERTa recall on neural text and average confidence RoBERTa predicted human-written with using various homoglyph pairs (with corresponding Unicode codes in parentheses). First row contains results on unaltered neural text. Unfortunately, default \LaTeX does not support all Unicode characters. However, a package was used to render the Cyrillic characters in this table. Examples of actual Cyrillic Unicode characters displayed can be seen in Appendix \ref{app_a} Figure \ref{homoglyph examples}.}
\label{first_homoglyph}
\begin{center}
\begin{tabular}{llll}
\multicolumn{1}{c}{\bf Original} &\multicolumn{1}{c}{\bf Homoglyph}  &\multicolumn{1}{c}{\bf Detector Recall} &\multicolumn{1}{c}{\bf Average Confidence}
\\ \hline \\
& &97.44\% &5.29\% \\
a (U+0061), e (U+0065) &\foreignlanguage{russian}{a} (U+0430), \foreignlanguage{russian}{e} (U+0435)    &13.57\%      &81.61\%     \\
e (U+0065) &\foreignlanguage{russian}{e}   (U+0435)        &16.11\% &79.43\% \\
e (U+0065) &é (U+00E9) &18.11\% &77.42\% \\
a (U+0061), c (U+0063) &\foreignlanguage{russian}{a} (U+0430), \foreignlanguage{russian}{c}  (U+0441)   &19.96\% &75.98\% \\

a (U+0061) &\foreignlanguage{russian}{a}  (U+0430)     &20.40\%  &75.55\%    \\
c (U+0063) & \foreignlanguage{russian}{c}  (U+0441)            &36.94\%  &61.78\%    \\
p (U+0070) &\foreignlanguage{russian}{p} (U+0440)          &42.25\% &56.99\%

\end{tabular}
\end{center}
\end{table}

\section{Results}
\label{others}

Results for the first homoglyph experiment can be seen in Table \ref{first_homoglyph}. Interestingly, replacing vowels with homoglyphs was a much more effective attack, even when the frequency of replacement was the same as that of consonants. Additionally, attacks using multiple homoglyph pairs were more effective than those which used only one.

For the second homoglyph experiment, according to Table \ref{first_homoglyph}, the most successful homoglyph pair was English ``e'' and English ``a'' to Cyrillic ``Ye'' and Cyrillic ``a'', respectively. When this homoglyph attack was allowed to replace all of the English ``e''s and English ``a''s in the neural text dataset, RoBERTa's recall on neural text dropped to 0.26\%.

\begin{figure}
    \includegraphics[scale=0.55]{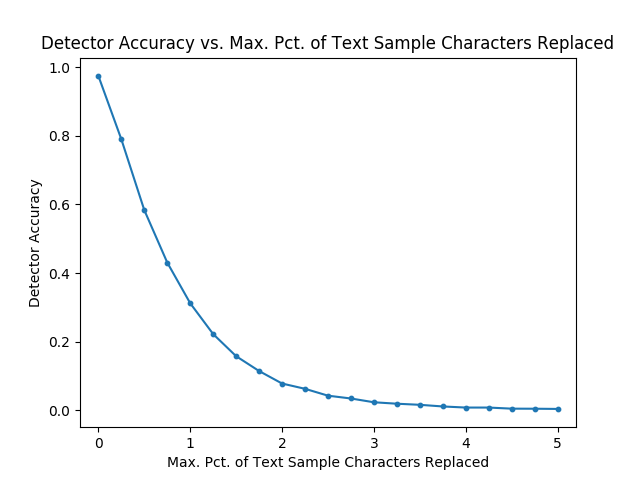}
    \caption{\centering RoBERTa neural text recall on neural text as a function of the maximum percentage of neural text sample characters a random homoglyph English ``e'' to Cyrillic ``Ye'' attack was allowed to replace.}
    \label{decrease}
\end{figure}

The results of the third homoglyph experiment can be seen in Figure \ref{decrease}. The most successful single character homoglyph attack was used. Neural text detector recall on neural text was inversely proportional to the amount of characters a homoglyph attack was allowed to replace.



The results of the fourth homoglyph experiment indicate that the homoglyph attacks are transferable to other neural text detectors. Before the English ``e'' and English ``a'' to Cyrillic ``Ye'' and Cyrillic ``a'' attack was implemented, GROVER predicted Machine++ for 19 of the 20 samples, and predicted Human++ for 1 of the 20 samples. After the homoglyph attack, GROVER predicted Machine++ for 3 of the 20 samples, Machine+ for 1 of the 20 samples, Human+ for 1 of the 20 samples, and Human++ for the remaining 15 samples. In an experiment testing the transferability of the homoglyph attack to GLTR, replacing all English ``e''s with Latin ``é''s across 10 neural text samples consistently shifted histograms and the way GLTR colored the given text in the online demo towards GLTR behavior characteristic of human writing. Graphical results can be seen in Appendix \ref{app_a}.

The results of the second misspelling experiment indicate that the misspelling attack is transferable to other neural text detectors as well. Before the misspelling attack was implemented, GROVER predicted Machine++ for 19 of the 20 samples, and predicted Human++ for 1 of the 20 samples. Note that random samples different from the ones used for the homoglyph transferability attack were used. After the homoglyph attack, GROVER predicted Machine++ for 8 of the 20 samples, Machine+ for 2 of the 20 samples, Human+ for 1 of the 20 samples, and Human++ for the remaining 9 samples. Similarly, the misspelling attack was also able to shift GLTR behavior towards that characteristic of humans across 10 neural text samples. An example of this can be seen in Appendix \ref{app_a}.

\section{Discussion}

It is interesting to note that the non-human like attacks were effective because they are not characteristic of human-written nor neural text, yet the neural text detectors predicted the text was human-written--just because the modified neural text wasn't characteristic of neural text. Clearly, automatic neural text detectors are trained not to discriminate between neural text and human-written text, but rather decide what is characteristic and uncharacteristic of neural text. As seen by the success of the homoglyph attacks presented in this paper, this creates a vulnerability for neural text detectors in which an adversary can change neural text to be characteristic of neither language models nor humans (e.g. mixing English and Cyrillic alphabets), yet have the modified neural text be classified as human-written text.

While homoglyph attacks may be defended against with similar tactics such as those employed by modern web-browsers and spell-check, human-like attacks will ultimately be much more difficult to defend against, especially as they increase in complexity and employ methods which create not just spelling errors, but also grammatical errors or different sampling mechanisms to encourage different word-choice. Such attacks will force neural text detectors to increasingly deepen their understanding of not only what constitutes neural text, but also what constitutes human-written text.

\section{Conclusion}

This work defines two classes of attacks on neural text detectors: non human-like and human-like. Both proved to be very effective in disrupting neural text detectors' ability to classify neural text accurately. Additionally, this paper sheds some light on what kinds of methods neural text detectors employ, and how these may be exploited. Future work should focus on making neural text detectors robust against the attacks presented in this work, and further explore the extent to which the attacks presented in this paper, particularly human-like attacks, may be deployed on neural text detectors.

\bibliography{iclr2020_conference}
\bibliographystyle{iclr2020_conference}

\appendix

\section{Shifting Neural Text's Distribution}
\label{shifting}

This experiment, similar to the ones performed by the GLTR authors, was designed to quantify the extent to which a homoglyph attack could shift the distribution of neural text away from that of text produced by a language model. The GPT-2 117M language model\footnote{An open-sourced GPT-2 117M model was taken from \url{https://github.com/huggingface/transformers}} was used to generate predictions for each token in a text sample. The token's position within GPT-2 117M's predictions, or rank, was then recorded. Lower ranks indicate an alignment with GPT-2 117M's predictions. 50 randomly chosen text samples from the WebText dataset made available in the same GitHub repository that provided the neural text dataset\footnote{\url{https://github.com/openai/gpt-2-output-dataset}} was used for human evaluation. 50 text samples were randomly chosen from the neural text dataset, which then had the English ``e" and English ``a'' to Cryllic ``Ye'' and Cyrillic ``a'' homoglyph attack with no maximum character replacement restriction applied to them. The results for the this experiment are displayed in Table \ref{ranks}. Overall, the homoglyph attack was successful in shifting neural text's distribution away from that of a language model.
\begin{table}[!htbp]
\caption{\centering Average rank of each word in GPT-2 117M predictions across 50 randomly chosen text samples for each category.}
\begin{center}
\begin{tabular}{l | l}
  &\multicolumn{1}{c}{\bf Rank}\\
 \hline 
\textbf{Human}   &128.26 \\ 
\textbf{Neural}   &14.05\\
\textbf{Homoglyph Neural}  &371.41

\end{tabular}
\end{center}
\label{ranks}
\end{table}

\newpage
\section{Examples}
\label{app_a}

\begin{figure}[!htbp]
    \centering
    \subfigure[]{{\includegraphics[width=12.5cm]{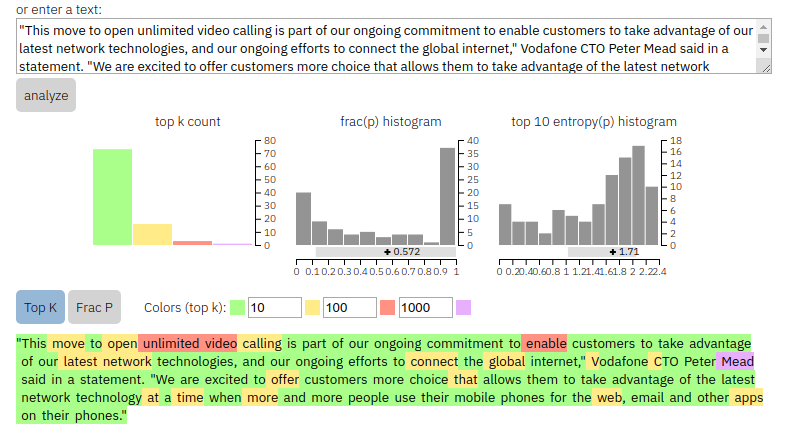}}}
    \subfigure[]{{\includegraphics[width=12.5cm]{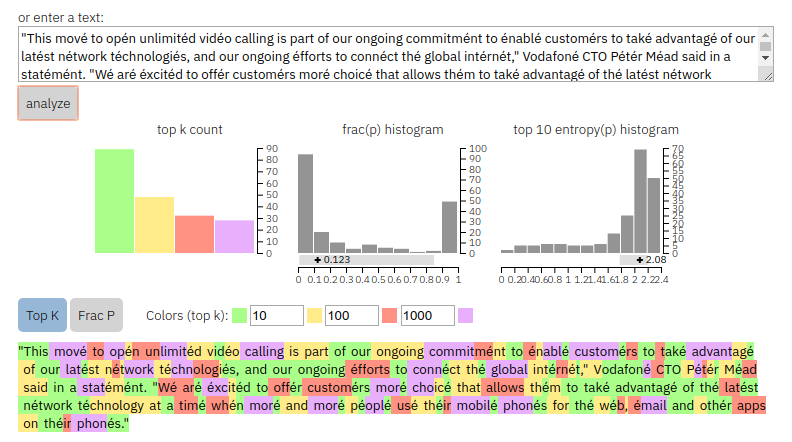}}}
    \caption{\centering{Top: GLTR output before homoglyph attack. Bottom: GLTR output after homoglyph attack. The presence of red and purple highlighted words indicates that GPT-2 117M had a difficult time predicting the word being highlighted, which helps human readers decide whether text was written by a language model or human.}}
    \label{fig:my_label}
\end{figure}

\begin{figure}[!htbp]
    \centering
    \subfigure[]{{\includegraphics[width=12.5cm]{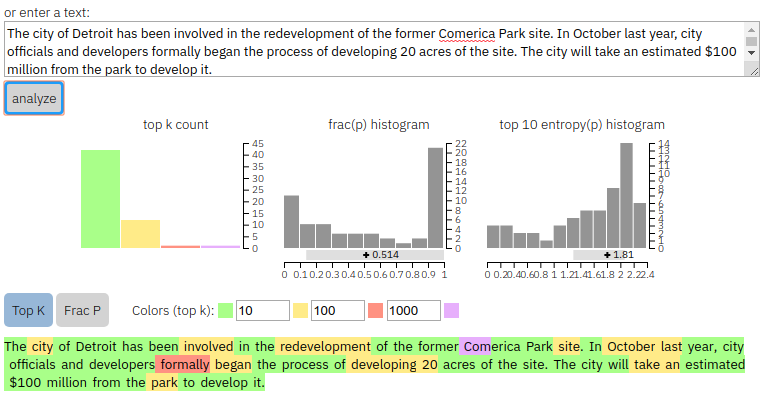}}}
    \subfigure[]{{\includegraphics[width=12.5cm]{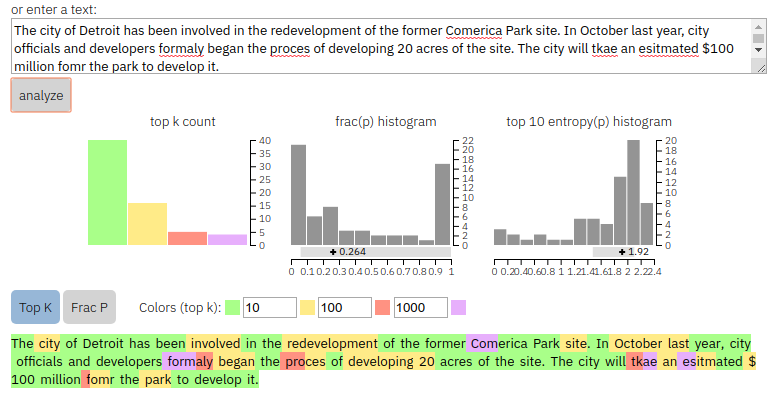}}}
    \caption{\centering{GLTR before (Subfigure (a)) and after (Subfigure (b)) the misspelling attack was applied.}}
    \label{fig:my_label}
\end{figure}

\begin{figure}[!htbp]
    \centering
    \subfigure[]{{\includegraphics[width=12.5cm]{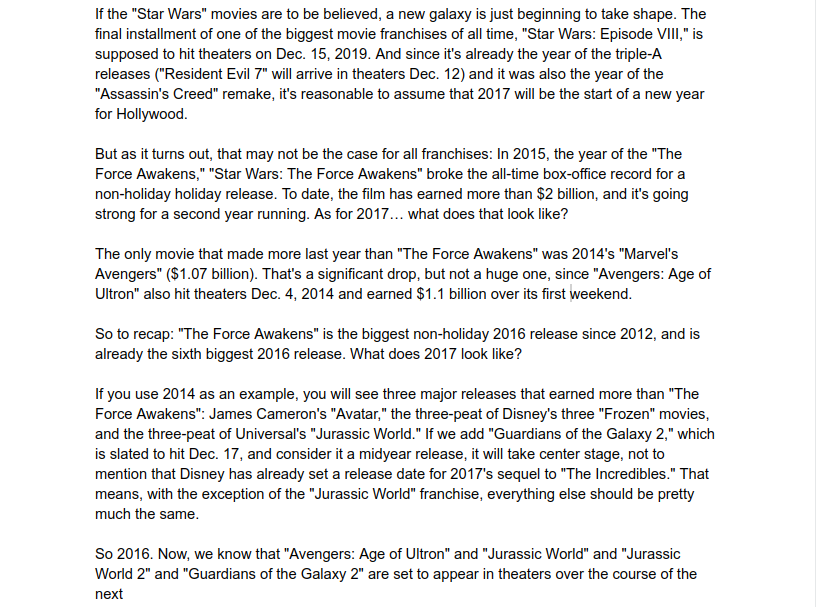}}}
    \subfigure[]{{\includegraphics[width=12.5cm]{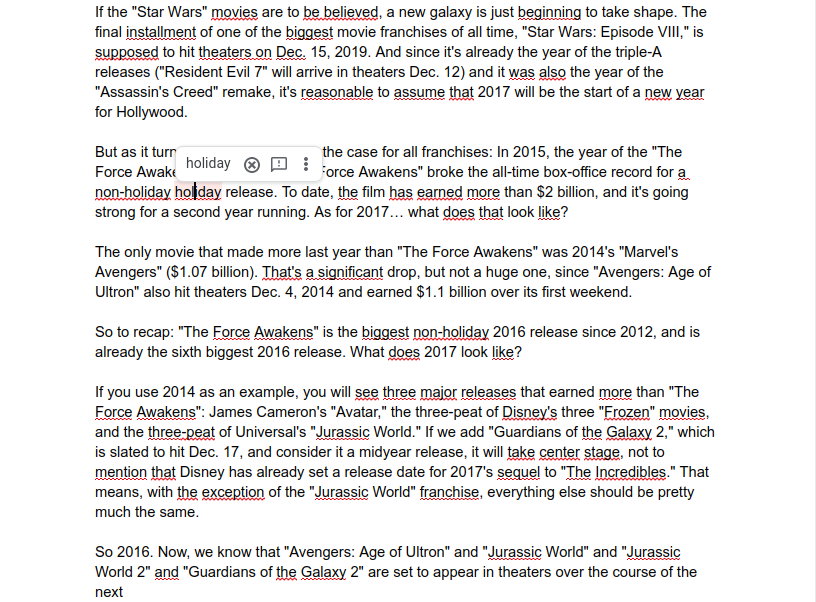}}}
    \caption{\centering The text in Subfigure (a) was classified as neural text by the neural text detector with 99.94\% confidence. Then, the English ``e'' and English ``a'' to Cyrillic ``Ye'' and Cyrillic ``a'' attack was applied to it. After this modification, the text sample, shown in Subfigure (b) was classified by RoBERTa as human-written with 98.50\% confidence. Both text samples are rendered in Google Docs. Interestingly, Google Docs's spell-correct program provides suggestions (marked by red underlining) to change SOME but not ALL of the words the attack modified back to their unaltered state.}
    \label{homoglyph examples}
\end{figure}

\end{document}